\documentclass[aps,floatfix,preprint,showpacs,preprintnumbers,nofootinbib,superscriptaddress,natbib]{revtex4}

\usepackage{graphicx,float}
\usepackage{epsfig}		
\usepackage{dcolumn}
\usepackage{bm}
\usepackage{amsmath,upgreek}
\usepackage{amssymb}
\usepackage{braket}
\usepackage{natbib}
\usepackage{letltxmacro} 

\usepackage[all]{xy}
\usepackage{pdfpages}
\usepackage{color}
\usepackage[colorlinks=true,
 linkcolor=red,
 urlcolor=purple,
 citecolor=blue,hyperindex]{hyperref}

\def\bkR{{\rm I\kern-.17em R}}

\def\0{\mbox{\tiny $0$}}
\def\1{\mbox{\tiny $1$}}
\def\2{\mbox{\tiny $2$}}
\def\3{\mbox{\tiny $3$}}
\def\4{\mbox{\tiny $4$}}
\def\5{\mbox{\tiny $5$}}
\def\6{\mbox{\tiny $6$}}
\def\7{\mbox{\tiny $7$}}
\def\8{\mbox{\tiny $8$}}
\def\9{\mbox{\tiny $9$}}

\def\f14{\mbox{\tiny $\frac{1}{4}$}}

\begin{document}

\title{Quantum cloning and teleportation fidelity in the noncommutative phase-space}

\author{P. Leal}
\email{up201002687@fc.up.pt}
\affiliation{Departamento de F\'isica e Astronomia and Centro de F\'isica do Porto, Faculdade de Ci\^{e}ncias da
Universidade do Porto, Rua do Campo Alegre 687, 4169-007, Porto, Portugal.}
\author{A. E. Bernardini}
\email{alexeb@ufscar.br}
\affiliation{Departamento de F\'isica e Astronomia and Centro de F\'isica do Porto, Faculdade de Ci\^{e}ncias da
Universidade do Porto, Rua do Campo Alegre 687, 4169-007, Porto, Portugal.}
\altaffiliation[On leave of absence from]{~Departamento de F\'{\i}sica, Universidade Federal de S\~ao Carlos, PO Box 676, 13565-905, S\~ao Carlos, SP, Brasil.}
\author{O. Bertolami}
\email{orfeu.bertolami@fc.up.pt}
\affiliation{Departamento de F\'isica e Astronomia and Centro de F\'isica do Porto, Faculdade de Ci\^{e}ncias da
Universidade do Porto, Rua do Campo Alegre 687, 4169-007, Porto, Portugal.}


\begin{abstract}
The formulation of the no-cloning theorem in the framework of phase-space noncommutative (NC) quantum mechanics (QM) is examined, and its implications for the computation of quantum cloning probabilities and teleportation fidelity are investigated through the Weyl-Wigner formulation of QM.
The principles of QM re-edited in terms of a deformed Heisenberg-Weyl algebra are shown to provide a covariant formulation for the quantum fidelity, through which the results from the no-cloning theorem for the ordinary QM can be reproduced.
Besides exhibiting an explicit correspondence between standard and NC QM for Gaussian Wigner functions, our results suggest a feasible interpretation for the NC continuous variable quantum cloning given in terms of quantum teleportation protocols.
\end{abstract}

\pacs{}
\keywords{noncommutativity, cloning, teleportation, phase-space}
\date{\today}
\maketitle

\maketitle

\section{Introduction}

The manipulation of quantum states for the processing of information has been stirred up by an increasing number of theoretical tools which describe the prospects of quantum-enhanced systems.
This includes several protocols for quantum factoring, quantum cloning and quantum teleportation, all of them involving multipartite quantum systems.
In the context of these procedures, a series of limitations and assumptions have been established by the so-called no-cloning theorem, which ensures that no random state can be duplicated \cite{Woot82}.
More specifically, the no-cloning theorem precludes the possibility of creating an auxiliar duplicate of a state during a quantum computation.
Fortunately, the advent of quantum error correcting codes \cite{Bern98,Brau04} allows for circumventing the limitations of the no-cloning hypothesis so to provide the setup for more engendered quantum computing protocols which involve, for instance, continuous variable framework \cite{Braunstein1998}.
Although originally designed for discrete variable quantum systems, the quantum platforms built from the continuous-spectrum described by the quadrature components of a light mode are presumably easier to handle than their discrete counterparts \cite{Cerf2000}.
In fact, quantum teleportation \cite{Braunstein1998} and quantum computation \cite{Braunstein1999} protocols, as well as quantum cryptographic schemes \cite{Mu1996} have been developed, all relying on continuous variables.

The aim of this paper is to investigate the foundations and the range of applicability of the no-cloning theorem in the framework of phase-space noncommutative (NC) quantum mechanics (QM).
Our analysis comprises a covariant formulation of quantifiers of the fidelity of states and a continuous variable quantum teleportation protocol in terms of NC variables.

NC extensions of QM and quantum cosmology have already been extensively examined \cite{Bastos3,Bastos3B,PedroLeal,Bastos,Gamboa,06A,08A,09A}.
The NC QM is suitably formulated in the Weyl-Wigner-Groenewold-Moyal (WWGM) formalism \cite{Groenewold,Moyal,Wigner} in a $2n$-dim phase-space where $n$ sets of canonical conjugate variables obey a deformed Heisenberg-Weyl (HW) algebra.
Our analysis will consist in revisiting the no-cloning and related theorems in QM, and generalize them for the deformed HW NC algebra \cite{Bastos,Gamboa,Rosenbaum} which supports the NC QM. Our study complements previous studies that examine NC features in quantum oscillations \cite{Rosenbaum,Prange,Nekrasov}, decoherence \cite{Bernardini13A,2015,Salomon}, quantum entanglement and locality \cite{Bernardini13B,Bernardini13B2}, uncertainty principle \cite{Bastos3,RSUP1,RSUP2,RSUP3}, equivalence principle \cite{Leal,Bastos4}, and gauge invariance \cite{Queiroz_2011,Leal}.

Our results clearly indicate that the no-cloning theorem can be generalized for continuous variables in the NC phase-space.
In fact, it will be shown that cloning and teleportation probabilities obtained from NC Wigner functions can be quantified through a NC covariant expression for the teleportation fidelity that allows for quantifying the reproducibility of quantum cloned and teleported Wigner functions, which fits perfectly the results for ordinary QM of Gaussian states.

The outline of this work is as follows.
In Sec.~II, the formulation of the NC QM as discussed in Ref.~\cite{Bastos} is reviewed.
In particular, one will be concerned with  the Wigner formulation of QM suitable for NC QM and the use of a generalized Seiberg-Witten (SW) map \cite{Seiberg} to define the NC Wigner function and some of its properties which will be relevant in the following sections.
The no-cloning theorem in the phase-space NC QM is obtained in Sec.~III.
Sec.~IV addresses the covariant formulation of the teleportation fidelity in the NC phase-space.
Since the cloning probabilities are also related to the entanglement fidelity, a relationship given in terms of NC Wigner functions is obtained and applied for quantifying the cloning probabilities of NC Gaussian states.
As will be seen, in the transition from ordinary QM to NC QM, the entanglement fidelity exhibits the same covariant behavior of the Wigner functions, a fundamental feature in demonstrating the no-cloning theorem in the NC framework.
Finally, in Sec.~V, a protocol for a teleportation process in phase-space is constructed so to provide an additional consistency test for reproducing quantum states in the NC framework.
Our concluding remarks, as drawn in Sec.~VI, do indicate that a consistent formulation of the NC version of the no-cloning theorem can be proven, so that its implications for teleportation protocols are preserved in the NC versions of the quantum information protocols.

\section{NC QM in phase-space}

The NC QM is characterized by the deformation of the HW algebra as represented by the commutation relations:
\begin{equation}
[\hat{x}_i, \hat{x}_j] = i\theta_{ij}, \quad [\hat{x}_i, \hat{p}_j] = i\hbar \delta_{ij}, \quad [\hat{p}_i, \hat{p}_j] = i\eta_{ij}, \quad i = 1,...n,
\label{ggreqs01}
\end{equation}
where $\theta_{ij}$ and $\eta_{ij}$ are skew-symmetric matrices with real entries, whereas in the ordinary QM one has
\begin{equation}
[\hat{q}_i, \hat{q}_j] = 0, \quad [\hat{q}_i, \hat{k}_j] = i\hbar \delta_{ij}, \quad [\hat{k}_i, \hat{k}_j] = 0, \quad i = 1,...n,
\label{ggreqs02}
\end{equation} 
with the variables of the two algebras being connected by the so-called SW map \cite{Seiberg},
\begin{equation}
\hat{x}_i = \lambda\hat{q}_i -\frac{\theta}{2\lambda\hbar}\epsilon_{ij} \hat{k}_j, \quad \hat{p}_i = \mu \hat{k}_i + \frac{\eta}{2\mu\hbar}\epsilon_{ij}\hat{q}_j,
\label{ggreqs03}
\end{equation}
which is invertible when the parameters $\lambda$ and $\mu$ are constrained by the relationship
\begin{equation}
{\theta \eta \over 4 \hbar^2} = \lambda \mu (\lambda \mu -1 ),
\label{constraint}
\end{equation}
so to give
\begin{eqnarray}
\mathit{\hat{q}}_i &=& \mu \left(1 - {\theta \eta\over\hbar^2} \right)^{- 1 / 2} \left( \hat{x}_i + {\theta\over 2 \lambda \mu \hbar} \epsilon_{ij} \hat{p}_j \right)~,\nonumber\\
\hat{k}_i &=& \lambda \left(1 - {\theta \eta\over\hbar^2} \right)^{-1 / 2} \left( \hat{p}_i-{\eta\over 2 \lambda \mu \hbar} \epsilon_{ij} \hat{x}_j \right),
\label{SWinverse}
\end{eqnarray}
with $\epsilon_{ij} = -\epsilon_{ji}$, $\theta\eta \lesssim \hbar^2$, and the corresponding Jacobian reading
\begin{equation}
{\partial (x,p)\over \partial (q,k)} = \sqrt{\det(\mbox{\boldmath$\Omega$})} = 1 - {\theta \eta\over\hbar^2}.
\label{Deter}\end{equation}

In its most general form, the SW map can be expressed as \cite{Bastos}:
\begin{equation}
\hat x_i = A_{ij} \hat q_j + B_{ij} \hat k_j, \hspace{1 cm}
\hat p_i = C_{ij} \hat q_j + D_{ij} \hat k_j,
\label{3.5}
\end{equation}
where ${\bf A},{\bf B},{\bf C},{\bf D}$ are real constant matrix solutions of the equations
\begin{equation}
{\bf A} {\bf D}^T - {\bf B} {\bf C}^T = {\bf I}_{n \times n} \hspace{1 cm} {\bf A} {\bf B}^T - {\bf B} {\bf A}^T = {1\over\hbar} {\bf \Theta} \hspace{1 cm}
{\bf C} {\bf D}^T - {\bf D} {\bf C}^T = {1\over\hbar} {\bf N}~,
\label{3.6}
\end{equation}
where the superscript ``$T$'' stands for matrix transposition and $A_{ij},\, B_{ij},\, C_{ij},\, D_{ij},\, \theta_{ij},\, \eta_{ij}$ are the entries of the matrices ${\bf A},\, {\bf B},\, {\bf C},\, {\bf D},\, {\bf \Theta},\, {\bf N}$, respectively. 

The above linear transformations imply that the NC algebra expressed by the relations from Eq.~(\ref{ggreqs01}) admits a representation in terms of the Hilbert space of the ordinary QM which provides a self-contained phase-space formulation of the NC QM. According to Ref.~\cite{Bastos},
the class of maps associated with the Weyl rule can be obtained by resorting to the ordinary Weyl-Wigner map in the following scheme (cf. Ref.~\cite{Bastos}),
\begin{equation}
{\underbrace{\begin{array}{ccccccc}
\hat a' (\hat {\bf z}) &\longrightarrow & \hat a(\hat{\mbox{\boldmath$\xi$}}) & \longrightarrow & a(\mbox{\boldmath$\xi$}) & \longrightarrow & a'({\bf z}) \\
& \hat{SW} & & W_{\xi} & & SW^{-1} &
\end{array}}_{W_z^{\xi}}},
\label{3.1.2}
\end{equation}
with $\mbox{\boldmath$\xi$}\equiv \{q_i,k_i\}$ and ${\bf z}\equiv \{x_i,p_i\}$.
Since the SW transformation is linear, there are no ordering ambiguities and thus $\hat{SW}$ and $SW$ have the same functional form.
Thus, the Wigner function $W_z^{\xi}$ provides the desired phase-space representation for NC QM if the NC properties are recast in the form of a new Moyal $\star$-product \cite{Moyal} in phase-space such that the full structure of the phase-space representation of NC QM is shown to be independent of the particular choice for the SW map \cite{Bastos}.

Considering the enveloping algebra of the extended Heisenberg algebra, $\hat{\cal A}_E({\cal H})$, and the algebra of Hilbert-Schmidt operators, $\hat{\cal A}({\cal H})$, it follows, from the SW map, that $\hat{\cal A}_E({\cal H}) =\hat{\cal A}({\cal H})$ \cite{Bastos}.
Using the generalized Weyl-Wigner map, from Eq.~(\ref{3.1.2}) in ${\cal A}(\bkR^{2n})$, for quantum operators such that $a,\, b \in {\cal A} (\bkR^{2n})$, the Moyal $\star$-product in the NC variables ${\bf z}=\{x_i,p_i\}$ reads
\begin{equation}
a({\bf z}) \star b({\bf z}) = a({\bf z}) e^{\frac{i \hbar}{2} \buildrel{\leftarrow}\over\partial_{z_{\alpha}}  \Omega_{\alpha \beta} \buildrel{\rightarrow}\over\partial_{z_{\beta}} }  b({\bf z}) = a({\bf z}) {\star}_{\hbar} {\star}_{\theta} {\star}_{\eta} b({\bf z}),
\label{3.1.6}
\end{equation}
where
\begin{eqnarray}
a({\bf z}) {\star}_{\hbar} b({\bf z}) &\equiv & a({\bf z})\, e^{\frac{i \hbar}{2} \buildrel{\leftarrow}\over\partial_{z_{i}} J_{ij}  \buildrel{\rightarrow}\over\partial_{z_{j}}} \,b({\bf z}), \label{3.1.7.A}\\
a({\bf z}) {\star}_{\theta} b({\bf z}) &\equiv&  a({\bf z})\, e^{\frac{i}{2} \frac{\buildrel{\leftarrow}\over\partial}{\partial x_i} \theta_{ij} \frac{\buildrel{\rightarrow}\over\partial}{\partial x_j}} \,b({\bf z}), \label{3.1.7.B}\\
a({\bf z}) {\star}_{\eta} b({\bf z}) &\equiv& a({\bf z})\, e^{\frac{i}{2} \frac{\buildrel{\leftarrow}\over\partial}{\partial p_i} \eta_{ij} \frac{\buildrel{\rightarrow}\over\partial}{\partial p_j}} \,b({\bf z}),
\label{3.1.7.C}
\end{eqnarray}
which can be cast into the compact notation,
\begin{equation}
a(\mbox{\boldmath$\chi$}) \star_{\Lambda} b(\mbox{\boldmath$\chi$}) = a(\mbox{\boldmath$\chi$}) e^{\frac{i}{2} \buildrel{\leftarrow}\over\partial_{\chi_r}  \Lambda_{rs} \buildrel{\rightarrow}\over\partial_{\chi_s} }  b(\mbox{\boldmath$\chi$})~,
\label{3.1.8}
\end{equation}
and where, for $\star$ and $\star_{\hbar}$, the variable $\mbox{\boldmath$\chi$}$ stands for $\mbox{\boldmath$\chi$}={\bf z}$ in the $2n$-dimensional phase-space, and the symplectic matrices in these cases read
\footnote{For $\star_{\theta}$ and $\star_{\eta}$, the variable $\mbox{\boldmath$\chi$}$ stands for $\mbox{\boldmath$\chi$}=\{x_i\}$ or $\mbox{\boldmath$\chi$}=\{p_i\}$ $(r,s=1, \cdots, n)$, and the symplectic matrices read
${\bf \Lambda} = {\bf \Theta}$, if $\star_{\Lambda} = \star_{\theta}$ and ${\bf \Lambda} = {\bf N}$, if $\star_{\Lambda} = \star_{\eta}$, respectively. According to Ref.~\cite{Bastos}, a $\star$-product of the form Eq.~(\ref{3.1.8}) acting on the space of polynomials on phase-space ($\star_{\Lambda} = \star$ or $\star_{\Lambda} = \star_{\hbar}$),
configuration space ($\star_{\Lambda} = \star_{\theta}$) or momentum space ($\star_{\Lambda} = \star_{\eta}$), can be represented as a Bopp shift,
\begin{equation}
a(\mbox{\boldmath$\chi$}) \star_{\Lambda} b(\mbox{\boldmath$\chi$}) = a \left(\mbox{\boldmath$\chi$} + \frac{i}{2} {\bf \Lambda} \buildrel{\rightarrow}\over\partial_{\chi} \right) b(\mbox{\boldmath$\chi$}) = a(\mbox{\boldmath$\chi$}) b \left(\mbox{\boldmath$\chi$} - \frac{i}{2} {\bf \Lambda} \buildrel{\leftarrow}\over\partial_{\chi} \right).
\label{3.1.11}
\end{equation}
}
\begin{equation}
\begin{array}{l l l l l}
{\bf \Lambda} = \hbar {\bf \Omega}, & \mbox{if } \star_{\Lambda} = \star & \hspace{0.5 cm} &,\hspace{0.5cm}
{\bf \Lambda} = \hbar {\bf J}, & \mbox{if } \star_{\Lambda} = \star_{\hbar}~,
\end{array}
\label{3.1.9}
\end{equation}
with
\begin{equation}
{\bf J} = \left(
\begin{array}{c c}
0 & {\bf I}_{n \times n}\\
- {\bf I}_{n \times n} &  0
\end{array}
\right)
\qquad
\mbox{and}
\qquad
{\bf \Omega} = \left(
\begin{array}{c c}
\frac{1}{\hbar} {\bf \Theta} & {\bf I}_{n \times n}\\
- {\bf I}_{n \times n} &  \frac{1}{\hbar} {\bf N}
\end{array}
\right).
\label{3.1.5}
\end{equation}
For ${\bf \Lambda}$ invertible, the $\star$-product of the form Eq.~(\ref{3.1.8}) admits a kernel representation:
\begin{equation}
a(\mbox{\boldmath$\chi$}) \star_{\Lambda} b(\mbox{\boldmath$\chi$}) = {1\over\pi^n | \det {\bf \Lambda} |} {\hspace{-.1cm}}\int{\hspace{-.15cm}} d \mbox{\boldmath$\chi$}' {\hspace{-.1cm}}\int{\hspace{-.15cm}} d \mbox{\boldmath$\chi$}'' \hspace{0.2 cm} a(\mbox{\boldmath$\chi$}') b(\mbox{\boldmath$\chi$}'')e^{[2i (\mbox{\boldmath$\chi$}- \mbox{\boldmath$\chi$}')^T {\bf \Lambda}^{-1} (\mbox{\boldmath$\chi$}''- \mbox{\boldmath$\chi$})]},
\label{3.1.14}
\end{equation}
for\footnote{Where $n$ stands for $2n$ or $n$ depending on whether the $\star$-product is $\star$, $\star_{\hbar}$, or $\star_{\theta}$, $\star_{\eta}$.} $a, b \in {\cal A}(\bkR^n)$.

To find the correspondence between ordinary and NC QM from the above results, one performs the SW transformation,
\begin{equation}
\mbox{\boldmath$\xi$} \longrightarrow {\bf z} = {\bf S} \mbox{\boldmath$\xi$} \hspace{0.2 cm},\hspace{0.2cm} \alpha(\mbox{\boldmath$\xi$}) \longrightarrow \alpha' ({\bf z}) = \alpha(\mbox{\boldmath$\xi$} ({\bf z})) \hspace{0.2 cm},\hspace{0.2cm} \beta(\mbox{\boldmath$\xi$}) \longrightarrow \beta' ({\bf z}) = \beta( \mbox{\boldmath$\xi$} ({\bf z})),
\label{3.1.20}
\end{equation}
with $S_{km} = \frac{\partial z_{k}}{\partial \xi_{m}}$, which reproduces the calculations for the corresponding ordinary variables, $\alpha(\mbox{\boldmath$\xi$} )$ and $\beta(\mbox{\boldmath$\xi$})$, namely the Weyl symbols for some operators $\hat \alpha$ and $\hat \beta$ in $\hat{{\cal A}} ({\cal H}) \cup \hat{{\cal F}}$, where $\hat{{\cal F}}$ are generic Hilbert-Schmidt operators,
such that the kernel representation of the Moyal $\star_{\hbar}$-product is given by \begin{equation}
\alpha(\mbox{\boldmath$\xi$}) \star_{\hbar} \beta( \mbox{\boldmath$\xi$}) = {1\over(\pi \hbar)^{2n}} {\hspace{-.1cm}}\int{\hspace{-.15cm}} d \mbox{\boldmath$\xi$}' {\hspace{-.1cm}}\int{\hspace{-.15cm}} d \mbox{\boldmath$\xi$}'' \hspace{0.2 cm} \alpha( \mbox{\boldmath$\xi$}') \beta(\mbox{\boldmath$\xi$}'') e^{[ - \frac{2i}{\hbar} (\mbox{\boldmath$\xi$} - \mbox{\boldmath$\xi$}')^T {\bf J} (\mbox{\boldmath$\xi$}'' - \mbox{\boldmath$\xi$})]}.
\label{3.1.19}
\end{equation}
Given that the symplectic matrix transforms as
${\bf \Omega} = {\bf S} {\bf J} {\bf S}^T
$, and, of course, $\det {\bf S} = \sqrt{\det {\bf \Omega}}$, under the SW map, Eq.~(\ref{3.1.19}) transforms according to \cite{Bastos}:
\begin{equation}
\alpha(\mbox{\boldmath$\xi$}) \star_{\hbar} \beta(\mbox{\boldmath$\xi$})\bigg{\vert}_{\mbox{\boldmath$\xi$}=\mbox{\boldmath$\xi$} ({\bf z})} ={1\over(\pi \hbar)^{2n}} {\hspace{-.1cm}}\int{\hspace{-.15cm}} d {\bf z}' {\hspace{-.1cm}}\int{\hspace{-.15cm}} d {\bf z}'' (\det {\bf S} )^{-2} \alpha' ({\bf z}') \beta' ({\bf z}'') e^{[ - \frac{2i}{\hbar} ({\bf z} - {\bf z}')^T ({\bf S}^{-1})^T {\bf J} {\bf S}^{-1} ({\bf z}'' - {\bf z})]},
\label{3.1.22}
\end{equation}
and, therefore,
\begin{equation}
\alpha' ({\bf z}) \star \beta' ({\bf z}) = {1\over(\pi \hbar)^{2n} |\det {\bf \Omega} |} {\hspace{-.1cm}}\int{\hspace{-.15cm}} d {\bf z}' {\hspace{-.1cm}}\int{\hspace{-.15cm}} d {\bf z}'' \hspace{0.2 cm} \alpha' ({\bf z}') \beta' ({\bf z}'') e^{[ \frac{2i}{\hbar} ({\bf z} - {\bf z}')^T {\bf \Omega}^{-1} ({\bf z}'' - {\bf z})]},
\label{3.1.23}
\end{equation}
through which, for $a({\bf z}) \equiv \alpha' ({\bf z})$ and $b({\bf z}) \equiv \beta' ({\bf z})$, one recovers Eq.~(\ref{3.1.14}).
From the above results, it follows that \cite{Bastos}

\vspace{0.3 cm}
{\noindent
\em {\bf Theorem:} For a Moyal $\star$-product of the form Eq.~(\ref{3.1.14}), one has
\begin{equation}
{\hspace{-.1cm}}\int{\hspace{-.15cm}} d \mbox{\boldmath$\chi$} \hspace{0.2 cm} A( \mbox{\boldmath$\chi$}) \star_{\Lambda} B( \mbox{\boldmath$\chi$}) = {\hspace{-.1cm}}\int{\hspace{-.15cm}} d \mbox{\boldmath$\chi$} \hspace{0.2 cm} A( \mbox{\boldmath$\chi$})  B( \mbox{\boldmath$\chi$}),
\label{3.1.25}
\end{equation}
similar to the WWGM formulation of ordinary QM.}

The proof is as follows.

{\noindent
{\bf Proof:} From the kernel representation, one has:
\begin{eqnarray}
{\hspace{-.1cm}}\int{\hspace{-.15cm}} {d \mbox{\boldmath$\chi$} A( \mbox{\boldmath$\chi$}) \star_{\Lambda} B( \mbox{\boldmath$\chi$})} &=& {1\over\pi^n | \det {\bf \Lambda} |} {\hspace{-.1cm}}\int{\hspace{-.15cm}} d \mbox{\boldmath$\chi$}  {\hspace{-.1cm}}\int{\hspace{-.15cm}} d \mbox{\boldmath$\chi$}' {\hspace{-.1cm}}\int{\hspace{-.15cm}} d  \mbox{\boldmath$\chi$}'' A( \mbox{\boldmath$\chi$}') B( \mbox{\boldmath$\chi$}'') e^{[2i ( \mbox{\boldmath$\chi$}- \mbox{\boldmath$\chi$}')^T {\bf \Lambda}^{-1} ( \mbox{\boldmath$\chi$}''- \mbox{\boldmath$\chi$})]}  \nonumber\\
&=& {1\over\pi^n | \det {\bf \Lambda} |}  {\hspace{-.1cm}}\int{\hspace{-.15cm}} d \mbox{\boldmath$\chi$}' {\hspace{-.1cm}}\int{\hspace{-.15cm}} d \mbox{\boldmath$\chi$}'' \pi^n | \det  {\bf \Lambda}| \delta( \mbox{\boldmath$\chi$}'-  \mbox{\boldmath$\chi$}'') A( \mbox{\boldmath$\chi$}') B( \mbox{\boldmath$\chi$}'') e^{ - 2i \mbox{\boldmath$\chi$}'^T {\bf \Lambda}^{-1}  \mbox{\boldmath$\chi$}''}\nonumber\\
&=& {\hspace{-.1cm}}\int{\hspace{-.15cm}} {d \mbox{\boldmath$\chi$} A( \mbox{\boldmath$\chi$})  B( \mbox{\boldmath$\chi$})}~,
\label{3.1.26}
\end{eqnarray}
where the antisymmetry of ${\bf \Lambda}$ has been used in the last step. 

From the above theorem, one can use the generalized Weyl-Wigner map to define the NC Wigner function.
For a system in a pure or mixed state represented by a density matrix $\hat{\rho} \in \hat{\cal F}$, the Wigner function in terms of the NC variables identified by ${\bf z}\equiv\{x_i,\,p_i\}$ is defined by \cite{Bastos}:
\begin{equation}
f^{NC}({\bf z}) \equiv  {1\over \sqrt{\det {\bf \Omega}}(2\pi\hbar)^n} W^{\xi}_z (\hat{\rho})~.
\label{3.3.1}
\end{equation}
It follows from the definition of $f^{NC}({\bf z})$ that if $\mbox{\boldmath$\xi$} \equiv \{q_i,\,k_i\}$ is the set of Heisenberg variables obtained via the SW map, the ordinary Wigner function associated with $\hat{\rho}$, $f^W (\mbox{\boldmath$\xi$}) \equiv \frac{1}{(2\pi\hbar)^n} W_{\xi} (\hat{\rho})$, is then given by
\begin{equation}
f^{NC}({\bf z})=\frac{1}{\sqrt{\det {\bf \Omega}}} f^W \left(\mbox{\boldmath$\xi$}({\bf z}) \right), \qquad \mbox{with}\quad {\hspace{-.1cm}}\int{\hspace{-.15cm}} dz \hspace{0.2 cm} f^{NC}({\bf z}) = 1,
\label{3.3.2}
\end{equation}
where $\mbox{\boldmath$\xi$}({\bf z})$ is the inverse transformation of the SW map.
Other properties of the NC Wigner function follow directly from application of the generalized Weyl-Wigner map \cite{Bastos}. Of course, in this section, only the results that are relevant for defining quantum cloning probabilities and teleportation fidelity in the NC phase-space have been addressed.

\section{No-cloning theorem in the NC phase-space} \label{section_Comm_NoClo}

In order to introduce the no-cloning theorem in the NC phase-space one will first review the well-known results of the no-cloning theorem in the Wigner formalism of phase-space QM.

First one considers two unknown states with Wigner functions $W_\psi(\mbox{\boldmath$\xi$})$ and $W_\phi(\mbox{\boldmath$\xi$})$ and, additionally, a blank third state, $W^e(\mbox{\boldmath$\xi$})$. All states are assumed to be normalized. Any cloning procedure should be able to take any unknown initial state (as well as an empty state) and replicate this state completely. In terms of QM, one should be able to find a unitary transformation such that,
\begin{equation}\label{out_state}
W^{out}_\psi(\mbox{\boldmath$\xi$}^A,\mbox{\boldmath$\xi$}^B)=U(\mbox{\boldmath$\xi$}^A,\mbox{\boldmath$\xi$}^B)\star\left[W_\psi(\mbox{\boldmath$\xi$}^A)W^e(\mbox{\boldmath$\xi$}^B)\right]\star U(\mbox{\boldmath$\xi$}^A,\mbox{\boldmath$\xi$}^B)
=W_\psi(\mbox{\boldmath$\xi$}^A)W_\psi(\mbox{\boldmath$\xi$}^B),
\end{equation}
for any state $\psi$. Here one uses the fact that for Hermitian operators, $U(\mbox{\boldmath$\xi$})^*=U(\mbox{\boldmath$\xi$})$. One should also notice that, for unitary transformations, $U(\mbox{\boldmath$\xi$}^A,\mbox{\boldmath$\xi$}^B)\star U(\mbox{\boldmath$\xi$}^A,\mbox{\boldmath$\xi$}^B)=1$. This also implies that for these transformations $U(\mbox{\boldmath$\xi$})^{-1}=U(\mbox{\boldmath$\xi$})^*$, where $U(\mbox{\boldmath$\xi$})^{-1}$ denotes the $\star$-product inverse. If one applies this procedure to the aforementioned states labeled by $\psi$ and $\phi$ one can then compute the integral:
\begin{equation}
\label{28}
P={\hspace{-.1cm}}\int{\hspace{-.15cm}} d\mbox{\boldmath$\xi$}^A{\hspace{-.1cm}}\int{\hspace{-.15cm}}d\mbox{\boldmath$\xi$}^B \,W^{out}_\psi(\mbox{\boldmath$\xi$}^A,\mbox{\boldmath$\xi$}^B)\,W^{out}_\phi(\mbox{\boldmath$\xi$}^A,\mbox{\boldmath$\xi$}^B).
\end{equation}
Indeed it can be carried out in two different ways, as illustrated in the following. First, by using the last equality on Eq.~\eqref{out_state} one gets:
\begin{equation}
\begin{split}
P=&{\hspace{-.1cm}}\int{\hspace{-.15cm}} d\mbox{\boldmath$\xi$}^A{\hspace{-.1cm}}\int{\hspace{-.15cm}}d\mbox{\boldmath$\xi$}^B\,W_\psi(\mbox{\boldmath$\xi$}^A)W_\psi(\mbox{\boldmath$\xi$}^B)\,W_\phi(\mbox{\boldmath$\xi$}^A)W_\phi(\mbox{\boldmath$\xi$}^B) \\
=&{\hspace{-.1cm}}\int{\hspace{-.15cm}} d\mbox{\boldmath$\xi$}^A\,W_\psi(\mbox{\boldmath$\xi$}^A)\,W_\phi(\mbox{\boldmath$\xi$}^A){\hspace{-.1cm}}\int{\hspace{-.15cm}} d\mbox{\boldmath$\xi$}^B\,W_\psi(\mbox{\boldmath$\xi$}^B)\,W_\phi(\mbox{\boldmath$\xi$}^B) \\
=&\left({\hspace{-.1cm}}\int{\hspace{-.15cm}} d\mbox{\boldmath$\xi$} \,W_\psi(\mbox{\boldmath$\xi$})\,W_\phi(\mbox{\boldmath$\xi$})\right)^2.
\end{split}
\end{equation}
However, if one uses the first equality of Eq.~\eqref{out_state} and the associativity of the $\star$-product, the integral is written as:
\begin{equation}
\begin{split}
P=&{\hspace{-.1cm}}\int{\hspace{-.15cm}} d\mbox{\boldmath$\xi$}^A{\hspace{-.1cm}}\int{\hspace{-.15cm}}d\mbox{\boldmath$\xi$}^B\,U(\mbox{\boldmath$\xi$}^A,\mbox{\boldmath$\xi$}^B)\star\left[W_\psi(\mbox{\boldmath$\xi$}^A)W^e(\mbox{\boldmath$\xi$}^B)\right]\star U(\mbox{\boldmath$\xi$}^A,\mbox{\boldmath$\xi$}^B) \\
&  U(\mbox{\boldmath$\xi$}^A,\mbox{\boldmath$\xi$}^B)\star\left[W_\phi(\mbox{\boldmath$\xi$}^A)W^e(\mbox{\boldmath$\xi$}^B)\right]\star U(\mbox{\boldmath$\xi$}^A,\mbox{\boldmath$\xi$}^B) \\
=& {\hspace{-.1cm}}\int{\hspace{-.15cm}} d\mbox{\boldmath$\xi$}^A{\hspace{-.1cm}}\int{\hspace{-.15cm}}d\mbox{\boldmath$\xi$}^B\,U(\mbox{\boldmath$\xi$}^A,\mbox{\boldmath$\xi$}^B)\star\left[W_\psi(\mbox{\boldmath$\xi$}^A)W^e(\mbox{\boldmath$\xi$}^B)\right]\star U(\mbox{\boldmath$\xi$}^A,\mbox{\boldmath$\xi$}^B) \star \\
& \star U(\mbox{\boldmath$\xi$}^A,\mbox{\boldmath$\xi$}^B)\star \left[W_\phi(\mbox{\boldmath$\xi$}^A)W^e(\mbox{\boldmath$\xi$}^B)\right]\star U(\mbox{\boldmath$\xi$}^A,\mbox{\boldmath$\xi$}^B) \\
=& {\hspace{-.1cm}}\int{\hspace{-.15cm}} d\mbox{\boldmath$\xi$}^A{\hspace{-.1cm}}\int{\hspace{-.15cm}}d\mbox{\boldmath$\xi$}^B\,U(\mbox{\boldmath$\xi$}^A,\mbox{\boldmath$\xi$}^B)\star\left[W_\psi(\mbox{\boldmath$\xi$}^A)W^e(\mbox{\boldmath$\xi$}^B)\right]\star  \left[W_\phi(\mbox{\boldmath$\xi$}^A)W^e(\mbox{\boldmath$\xi$}^B)\right]\star U(\mbox{\boldmath$\xi$}^A,\mbox{\boldmath$\xi$}^B), \\
\end{split}
\end{equation}
where one has used the identity (cf. Eq. (\ref{3.1.25})):
\begin{equation} \label{star_integral}
{\hspace{-.1cm}}\int{\hspace{-.15cm}} d\mbox{\boldmath$\xi$}\,A(\mbox{\boldmath$\xi$})\star B(\mbox{\boldmath$\xi$})={\hspace{-.1cm}}\int{\hspace{-.15cm}} d\mbox{\boldmath$\xi$}\,A(\mbox{\boldmath$\xi$})\,B(\mbox{\boldmath$\xi$}),
\end{equation}
and the unitarity of $U(\mbox{\boldmath$\xi$}^A,\mbox{\boldmath$\xi$}^B)$.
By following the $\star$-product associativity and the commutativity of the regular product, one obtains:
\begin{equation}
\begin{split}
P =& {\hspace{-.1cm}}\int{\hspace{-.15cm}} d\mbox{\boldmath$\xi$}^A{\hspace{-.1cm}}\int{\hspace{-.15cm}}d\mbox{\boldmath$\xi$}^B\,\left\{U(\mbox{\boldmath$\xi$}^A,\mbox{\boldmath$\xi$}^B)\star\left[W_\psi(\mbox{\boldmath$\xi$}^A)W^e(\mbox{\boldmath$\xi$}^B)\right]\star \left[W_\phi(\mbox{\boldmath$\xi$}^A)W^e(\mbox{\boldmath$\xi$}^B)\right]\right\} U(\mbox{\boldmath$\xi$}^A,\mbox{\boldmath$\xi$}^B) \\
=& {\hspace{-.1cm}}\int{\hspace{-.15cm}} d\mbox{\boldmath$\xi$}^A{\hspace{-.1cm}}\int{\hspace{-.15cm}}d\mbox{\boldmath$\xi$}^B\,U(\mbox{\boldmath$\xi$}^A,\mbox{\boldmath$\xi$}^B)\star U(\mbox{\boldmath$\xi$}^A,\mbox{\boldmath$\xi$}^B)\star\left[W_\psi(\mbox{\boldmath$\xi$}^A)W^e(\mbox{\boldmath$\xi$}^B)\right]\star \left[W_\phi(\mbox{\boldmath$\xi$}^A)W^e(\mbox{\boldmath$\xi$}^B)\right] \\
=& {\hspace{-.1cm}}\int{\hspace{-.15cm}} d\mbox{\boldmath$\xi$}^A{\hspace{-.1cm}}\int{\hspace{-.15cm}}d\mbox{\boldmath$\xi$}^B\,\left[W_\psi(\mbox{\boldmath$\xi$}^A)W^e(\mbox{\boldmath$\xi$}^B)\right]\star\left[W_\phi(\mbox{\boldmath$\xi$}^A)W^e(\mbox{\boldmath$\xi$}^B)\right],
\end{split}
\end{equation}
which can be recast in the simplified form of:
\begin{equation}
\begin{split}
P=& {\hspace{-.1cm}}\int{\hspace{-.15cm}} d\mbox{\boldmath$\xi$}^A{\hspace{-.1cm}}\int{\hspace{-.15cm}}d\mbox{\boldmath$\xi$}^B\,\left[W_\psi(\mbox{\boldmath$\xi$}^A)W^e(\mbox{\boldmath$\xi$}^B)\right]\left[W_\phi(\mbox{\boldmath$\xi$}^A)W^e(\mbox{\boldmath$\xi$}^B)\right] \\
=&{\hspace{-.1cm}}\int{\hspace{-.15cm}} d\mbox{\boldmath$\xi$}^A\,W_\psi(\mbox{\boldmath$\xi$}^A)\,W_\phi(\mbox{\boldmath$\xi$}^A){\hspace{-.1cm}}\int{\hspace{-.15cm}} d\mbox{\boldmath$\xi$}^B\,W^e(\mbox{\boldmath$\xi$}^B)^2 \\
=&{\hspace{-.1cm}}\int{\hspace{-.15cm}} d\mbox{\boldmath$\xi$}\,W_\psi(\mbox{\boldmath$\xi$})\,W_\phi(\mbox{\boldmath$\xi$}).
\end{split}
\end{equation}

Therefore, this cloning procedure implies that $P^2=P$. Since $P$ is a real number, then the only possible solutions are that $P=0$ or $P=1$. This in turn implies that either the states are orthogonal or are the same state, which is absurd since it was assumed that the associated  $\psi$ and $\phi$ states were generic and unknown. This leads to the conclusion that such a cloning mechanism does not exist, proving the no-cloning theorem.

Most importantly, having the above derivation in mind, one should notice that the properties used there are only concerned with the $\star$-product defined for the theory. Thus, according to the results from Ref.~\cite{Bastos} expressed by theorem Eq.~(\ref{star_integral}), and the ensued invariance of Eq.~(\ref{28}) under the SW map, and by the associativity of the NC $\star_{NC}$-product, a proof as the above one can be extended to NC QM. In fact, this generalizes the no-cloning theorem to any theory where the $\star$-product obeys the two aforementioned properties: $\star_{\Lambda}$-product associativity and the theorem, Eq.~(\ref{star_integral})\footnote{Some of the steps of this result have already been obtained in Ref.~\cite{Bruno}, although through a different procedure, which did not rely on the phase-space Wigner formalism for NC QM. The method employed here allows for the generalization to any deformation of the HW algebra.}.

A similar argument can also be used to prove the converse of the no-cloning theorem: the no-deleting theorem. 
In this case, one should look for a unitary transformation that yields:
\begin{equation}\label{out_state_2}
\begin{split}
W^{out}_\psi(\mbox{\boldmath$\xi$}^A,\mbox{\boldmath$\xi$}^B)&=U(\mbox{\boldmath$\xi$}^A,\mbox{\boldmath$\xi$}^B)\star\left[W_\psi(\mbox{\boldmath$\xi$}^A)W_\psi(\mbox{\boldmath$\xi$}^B)\right]\star U(\mbox{\boldmath$\xi$}^A,\mbox{\boldmath$\xi$}^B)= \\
&=W_\psi(\mbox{\boldmath$\xi$}^A)W^e(\mbox{\boldmath$\xi$}^B).
\end{split}
\end{equation}
Considering two states, $\psi$ and $\phi$, and using the second equality we get:
\begin{equation}
\begin{split}
P=&{\hspace{-.1cm}}\int{\hspace{-.15cm}} d\mbox{\boldmath$\xi$}^A{\hspace{-.1cm}}\int{\hspace{-.15cm}}d\mbox{\boldmath$\xi$}^B\,W_\psi(\mbox{\boldmath$\xi$}^A)W^e(\mbox{\boldmath$\xi$}^B)\,W_\phi(\mbox{\boldmath$\xi$}^A)W^e(\mbox{\boldmath$\xi$}^B) \\
=&{\hspace{-.1cm}}\int{\hspace{-.15cm}} d\mbox{\boldmath$\xi$}^A\,W_\psi(\mbox{\boldmath$\xi$}^A)\,W_\phi(\mbox{\boldmath$\xi$}^A){\hspace{-.1cm}}\int{\hspace{-.15cm}} d\mbox{\boldmath$\xi$}^B\,W^e(\mbox{\boldmath$\xi$}^B)^2\\
=&\int{\hspace{-.15cm}} d\mbox{\boldmath$\xi$} \,W_\psi(\mbox{\boldmath$\xi$})\,W_\phi(\mbox{\boldmath$\xi$}).
\end{split}
\end{equation}
However, if instead, the first equality is used, after some straightforward algebra, the result becomes:
\begin{equation}
\begin{split}
P=&{\hspace{-.1cm}}\int{\hspace{-.15cm}} d\mbox{\boldmath$\xi$}^A{\hspace{-.1cm}}\int{\hspace{-.15cm}}d\mbox{\boldmath$\xi$}^B\,U(\mbox{\boldmath$\xi$}^A,\mbox{\boldmath$\xi$}^B)\star\left[W_\psi(\mbox{\boldmath$\xi$}^A)W_\psi(\mbox{\boldmath$\xi$}^B)\right]\star U(\mbox{\boldmath$\xi$}^A,\mbox{\boldmath$\xi$}^B) \\
&  U(\mbox{\boldmath$\xi$}^A,\mbox{\boldmath$\xi$}^B)\star\left[W_\phi(\mbox{\boldmath$\xi$}^A)W_\phi(\mbox{\boldmath$\xi$}^B)\right]\star U(\mbox{\boldmath$\xi$}^A,\mbox{\boldmath$\xi$}^B) \\
=& {\hspace{-.1cm}}\int{\hspace{-.15cm}} d\mbox{\boldmath$\xi$}^A{\hspace{-.1cm}}\int{\hspace{-.15cm}}d\mbox{\boldmath$\xi$}^B\,U(\mbox{\boldmath$\xi$}^A,\mbox{\boldmath$\xi$}^B)\star\left[W_\psi(\mbox{\boldmath$\xi$}^A)W_\psi(\mbox{\boldmath$\xi$}^B)\right]\star U(\mbox{\boldmath$\xi$}^A,\mbox{\boldmath$\xi$}^B) \star \\
& \star U(\mbox{\boldmath$\xi$}^A,\mbox{\boldmath$\xi$}^B)\star \left[W_\phi(\mbox{\boldmath$\xi$}^A)W_\phi(\mbox{\boldmath$\xi$}^B)\right]\star U(\mbox{\boldmath$\xi$}^A,\mbox{\boldmath$\xi$}^B) \\
=& {\hspace{-.1cm}}\int{\hspace{-.15cm}} d\mbox{\boldmath$\xi$}^A{\hspace{-.1cm}}\int{\hspace{-.15cm}}d\mbox{\boldmath$\xi$}^B\,U(\mbox{\boldmath$\xi$}^A,\mbox{\boldmath$\xi$}^B)\star\left[W_\psi(\mbox{\boldmath$\xi$}^A)W_\phi(\mbox{\boldmath$\xi$}^B)\right]\star \left[W_\phi(\mbox{\boldmath$\xi$}^A)W_\phi(\mbox{\boldmath$\xi$}^B)\right]\star U(\mbox{\boldmath$\xi$}^A,\mbox{\boldmath$\xi$}^B) \\
=& {\hspace{-.1cm}}\int{\hspace{-.15cm}} d\mbox{\boldmath$\xi$}^A{\hspace{-.1cm}}\int{\hspace{-.15cm}}d\mbox{\boldmath$\xi$}^B\,\left\{U(\mbox{\boldmath$\xi$}^A,\mbox{\boldmath$\xi$}^B)\star\left[W_\psi(\mbox{\boldmath$\xi$}^A)W_\psi(\mbox{\boldmath$\xi$}^B)\right] \star \left[W_\phi(\mbox{\boldmath$\xi$}^A)W_\phi(\mbox{\boldmath$\xi$}^B)\right]\right\} U(\mbox{\boldmath$\xi$}^A,\mbox{\boldmath$\xi$}^B) \\
=& {\hspace{-.1cm}}\int{\hspace{-.15cm}} d\mbox{\boldmath$\xi$}^A{\hspace{-.1cm}}\int{\hspace{-.15cm}}d\mbox{\boldmath$\xi$}^B\,U(\mbox{\boldmath$\xi$}^A,\mbox{\boldmath$\xi$}^B)\star U(\mbox{\boldmath$\xi$}^A,\mbox{\boldmath$\xi$}^B)\star\left[W_\psi(\mbox{\boldmath$\xi$}^A)W_\psi(\mbox{\boldmath$\xi$}^B)\right]\star \left[W_\phi(\mbox{\boldmath$\xi$}^A)W_\phi(\mbox{\boldmath$\xi$}^B)\right] \\
=& {\hspace{-.1cm}}\int{\hspace{-.15cm}} d\mbox{\boldmath$\xi$}^A{\hspace{-.1cm}}\int{\hspace{-.15cm}}d\mbox{\boldmath$\xi$}^B\,\left[W_\psi(\mbox{\boldmath$\xi$}^A)W_\psi(\mbox{\boldmath$\xi$}^B)\right]\star\left[W_\phi(\mbox{\boldmath$\xi$}^A)W_\phi(\mbox{\boldmath$\xi$}^B)\right] \\
=& {\hspace{-.1cm}}\int{\hspace{-.15cm}} d\mbox{\boldmath$\xi$}^A{\hspace{-.1cm}}\int{\hspace{-.15cm}}d\mbox{\boldmath$\xi$}^B\,\left[W_\psi(\mbox{\boldmath$\xi$}^A)W_\psi(\mbox{\boldmath$\xi$}^B)\right]\left[W_\phi(\mbox{\boldmath$\xi$}^A)W_\phi(\mbox{\boldmath$\xi$}^B)\right] \\
=&{\hspace{-.1cm}}\int{\hspace{-.15cm}} d\mbox{\boldmath$\xi$}^A\,W_\psi(\mbox{\boldmath$\xi$}^A)\,W_\phi(\mbox{\boldmath$\xi$}^A){\hspace{-.1cm}}\int{\hspace{-.15cm}} d\mbox{\boldmath$\xi$}^B\,W_\psi(\mbox{\boldmath$\xi$}^B)\,W_\phi(\mbox{\boldmath$\xi$}^B) \\
=&\left({\hspace{-.1cm}}\int{\hspace{-.15cm}} d\mbox{\boldmath$\xi$}\,W_\psi(\mbox{\boldmath$\xi$})\,W_\phi(\mbox{\boldmath$\xi$})\right)^2.
\end{split}
\end{equation}
Thus, one concludes that $P=P^2$, which, by the same argument used on the QM no-cloning theorem, proves that there is no unitary transformation that acts on states as specified in Eq. (\ref{out_state_2}). Thus, the no-deleting theorem follows. Again, this is valid for any deformation of the HW algebra that gives rise to a $\star_{\Lambda}$-product that is associative and obeys Eq. (\ref{star_integral}), namely in a particular framework of the NC QM.

\section{Teleportation fidelity in the NC phase-space}

The capability for teleporting quantum information is associated to the irreducible nonlocal content of QM, which is exemplified by the nonlocal features of an entangled quantum state \cite{Braunstein2000}.
Given the entanglement dynamics of two interacting subsystems of a composite quantum system \cite{EPR}, the teleportation of a single-mode quantum state can be engendered through a suitable variation of the original Einstein-Podolsky-Rosen (EPR) procedure \cite{EPR,Braunstein1998}. 
Once expressed in terms of continuous variable systems, the original formulation of the procedure brings about the nonlocal properties shared by two subsystems in the EPR state with perfect correlations in both position and momentum coordinates.
These correlations can be expressed in the language of bipartite Gaussian states \cite{Buono1} for which the corresponding EPR phase-space Wigner function is written as \cite{Bernardini13B}:
\begin{equation}
\label{Wigner1}
W_{EPR}({\bf{z}})=\frac{1}{\pi^2{\sqrt{\det[{\mbox{\boldmath$\Sigma$}}_2]}}}{\exp{\left(-{\bf{z}}^{T}{\mbox{\boldmath$\Sigma$}}_2^{-1}{\bf{z}}\right)}}~,
\end{equation}
where, in this case, ${\bf z}\equiv(x_1,\,p_1,\,x_2,\,p_2)$, which is associated to a set of orthogonal quadratures, for modes $\alpha_1 \equiv x_1 +i\,p_1$ and $\alpha_2\equiv x_2 +i\,p_2$, and the covariance matrix, ${\mbox{\boldmath$\Sigma$}}_2$, given by 
\begin{equation}
\label{CM1}
{\mbox{\boldmath$\Sigma$}}_2=\frac{1}{2}\left(
\begin{array}{c c}
\beta & \gamma \\
\gamma^{T} & \beta
\end{array}\right),
\end{equation}
where $\beta = \cosh(2r) \,\mathrm{Diag}[+1\,+1]$ represents the self-correlation of single subsystems and $\gamma= -\sinh(2r)\sigma^z_{(2)} = \sinh(2r) \,\mathrm{Diag}[-1\,+1]$ describes the correlation between the two subsystems, with both given in terms of the associated squeezing parameter, $r$.

Here, the real vector ${\bf z}$ defines the set of canonically conjugate variables, position and momentum, for the relevant pathways for a massive particle and quadrature amplitudes suitably associated to electromagnetic field modes. This yields:
\begin{equation}
\label{Wigner2}
W_{EPR}(\alpha_1,\,\alpha_2)=\frac{4}{\pi^2}{\exp{\left[-\cosh(2r)\left(\vert\alpha_1\vert^2+\vert\alpha_2\vert^2\right)-2\sinh(2r)\,\mathrm{Re}[\alpha_1\alpha_2]\right]}}.
\end{equation}

The entangled state, $W_{EPR}$, works as an auxiliary tool shared by {\em input} and {\em output} states, $\hat{\rho}_{in}$ and $\hat{\rho}_{out}$, for the construction of realistic teleportation protocols described by continuous variable quantum states \cite{Braunstein1998,Braunstein2000}. These are the basic engineering tools of teleportation protocols \cite{Aux1,Aux2,Aux3}. These protocols give support to the development of the convolutional formalism that will be considered in the construction of a NC version of a fidelity quantifier for teleported quantum states \cite{Schumacher1996}.

Turning the notation to the above mentioned single-mode continuous variable, $\alpha$, the framework  sets up the teleportation evolution described by \cite{Braunstein2000}
\begin{equation}
\hat{\rho}_{out} = {\hspace{-.1cm}}\int{\hspace{-.15cm}} d\alpha{\hspace{-.1cm}}\int{\hspace{-.15cm}} d\alpha^* \, \hat{D}(\alpha) \, \hat{\rho}_{in}\, \hat{D}^{\dagger}(\alpha),\label{009}
\end{equation}
where $\hat{\rho}_{in}$ is the original teleported state and $\hat{D}(\alpha) = \exp\left(\alpha a^{\dagger} - \alpha^* a\right)$ is the displacement operator \cite{Aux1,Aux2,Cerf2000} such that $\hat{D}(\alpha) \vert\omega\rangle = \vert\omega + \alpha\rangle$.
The quantitative measure of the reproducibility of the {\em input} state is prescribed by an {\em output} state, $\hat{\rho}_{out}$ (c.f. Eq.~(\ref{009})), given by means of the entanglement fidelity \cite{Schumacher1996}:
\begin{equation}
\mathcal{F}_{E} = {\hspace{-.1cm}}\int{\hspace{-.15cm}} d\alpha{\hspace{-.1cm}}\int{\hspace{-.15cm}} d\alpha^* \, \hat{D}(\alpha) \, \bigg{\vert} Tr\left[\hat{D}(\alpha)\,\hat{\rho}_{in}\right]\bigg{\vert}^2\label{011},
\end{equation}
which, in the context of phase-space variables, can be reconstructed in terms of Wigner functions.

The overall analysis, which includes finite (nonsingular) degrees of correlation and incorporate inefficiencies in the measurement process \cite{Braunstein1998}, results into a continuous variable cloning protocol translated into a convolutional formalism \cite{convolutional} statistically expressed by the correspondence between {\em input} and {\em output} states in terms of Wigner functions of quantum ensembles, $W_{in}$ and $W_{out}$, related by
\begin{equation}
W_{out} = W_{in} \circ G_{\sigma},
\label{008}
\end{equation}
where $G_{\sigma} = \pi^{-2}\,\exp(-\vert\alpha\vert^{\2}/\sigma)$ with $\sigma =\exp(-2r)$, and which of course can be extended to a multimodal framework such that $\sigma \to {\mbox{\boldmath$\Sigma$}}$, the covariance matrix.

According to the convolutional relation, Eq.~\eqref{008}, the entanglement fidelity between {\em input} and {\em output} $n$-modal Wigner functions can be written as
\begin{equation}
\mathcal{F}^W_{\Sigma} = (2\pi)^n{\hspace{-.1cm}}\int{\hspace{-.15cm}} d^{n}{\bf z}\, W_{out}({\bf z}) W_{in}({\bf z}) =(2\pi)^n\, {\hspace{-.1cm}}\int{\hspace{-.15cm}} d^{n}{\bf z}{\hspace{-.1cm}}\int{\hspace{-.15cm}} d^{n}{\bf z}'\, W_{in}({\bf z}')\,G_{\Sigma}({\bf z} - {\bf z}')\,W_{in}({\bf z}) ,
\label{012}
\end{equation}
with ${\bf z}\equiv(x_1,\,p_1,\,x_2,\,p_2,\,x_3,\,p_3,\,\dots,\,x_n,\,p_n,)$, and
\begin{equation}
W_{out}({\bf z}) = {\hspace{-.1cm}}\int{\hspace{-.15cm}} d^{n}{\bf z}'\, W_{in}({\bf z}')\,G_{\Sigma}({\bf z} - {\bf z}'),
\label{012B}
\end{equation}
where, in this case, the Gaussian cloner is given by
\begin{equation}
G_{\Sigma}({\bf z}) = \frac{1}{\pi^{n}{\sqrt{Det[{\mbox{\boldmath$\Sigma$}}]}}}{\exp{\left(-{\bf{z}}^{T}{\mbox{\boldmath$\Sigma$}}^{-1}{\bf{z}}\right)}},
\label{012C}
\end{equation}
for an arbitrary {\mbox{\boldmath$\Sigma$}}.
For normalized {\em input} Wigner functions,
\begin{equation}
 {\hspace{-.1cm}}\int{\hspace{-.15cm}} d^{n}{\bf z}\, W_{in}({\bf z}) = 1,
\label{013}
\end{equation}
and a $n$-partite Gaussian distribution, $W_{in}({\bf z}) = G_{\Sigma}({\bf z})$, one obtains the maximal value of the entanglement fidelity, which is given by $2^n/(3^n{\sqrt{Det(\mbox{\boldmath$\Sigma$})}})$.
For pure states, with ${\sqrt{Det(\mbox{\boldmath$\Sigma$})}}=1$, single-mode Gaussians exhibit optimal cloning fidelity, which leads to duplications of coherent Gaussian states with a fidelity of $2/3$ \cite{Cerf2000}.
A unitary cloning transformation identified by the above Wigner function teleportation protocol provides two copies of a system with a continuous spectrum at the price of a non unity cloning fidelity where the cloner is identified by
\begin{equation}
\lim_{\Sigma\to\infty} G_{\Sigma}({\bf z} - {\bf z}') = \delta^{(n)}({\bf z} - {\bf z}'),
\label{012CC}
\end{equation}
which, in this case, is reached by the implementation of an uncertainty principle (UP) violating protocol which returns 
\begin{equation}
\lim_{\Sigma\to\infty} \mathcal{F}^W_{\Sigma} = (2\pi)^n\, {\hspace{-.1cm}}\int{\hspace{-.15cm}} d^{n}{\bf z} W^2_{in}({\bf z}),\label{014}
\end{equation}
that is, the purity of the quantum state $W_{in}({\bf z})$. For $W_{in}({\bf z})$ identified by a Gaussian $G_{\Gamma}({\bf z})$, one has
\begin{equation}
\lim_{\Sigma\to\infty} \mathcal{F}^\Gamma_{\Sigma} = (2\pi)^n\, {\hspace{-.1cm}}\int{\hspace{-.15cm}} d^{n}{\bf z} W^2_{in}({\bf z}) = 
(2\pi)^n \frac{1}{(2\pi)^n{\sqrt{Det[{\mbox{\boldmath$\Gamma$}}]}}} ={ 1\over{\sqrt{Det[{\mbox{\boldmath$\Gamma$}}]}}},
\label{014B}
\end{equation}
the Gaussian purity.

In fact, the result from Eq.~(\ref{012}) for the hypothesis, Eq.~(\ref{012C}), can be straightforwardly generalized to Gaussian states given by $G_{\Gamma}({\bf z})$. In this case, $W_{in}({\bf z}) = G_{\Sigma}({\bf z})$, and with some involved mathematical manipulations, leads to \cite{Math}:
\begin{eqnarray}
\mathcal{F}^\Gamma_{\Sigma} &=& (2\pi)^n\, {\hspace{-.1cm}}\int{\hspace{-.15cm}} d^{n}{\bf z}{\hspace{-.1cm}}\int{\hspace{-.15cm}} d^{n}{\bf z}'\, G_{\Gamma}({\bf z}')\,G_{\Sigma}({\bf z} - {\bf z}')\,G_{\Gamma}({\bf z}) = \frac{2^n}{\sqrt{{Det[2{\mbox{\boldmath$\Gamma$}}+\mbox{{\boldmath$\Sigma$}}]}}}.
\label{012DD}
\end{eqnarray}
In the context of the NC QM, for the case where ${\bf z}$ is a vector in the NC phase-space, which is related to the commutative one by the linear SW map given by ${\bf z} = {\bf S} \mbox{\boldmath$\xi$}$, a quite relevant result is that the covariance of a NC fidelity quantifier, $\mathcal{F}^W_{\Sigma} \to \mathcal{F}^{NC}_{\Sigma}$, under the SW map, i.e. an analogous quantifier in the commutative phase-space, where $\mathcal{F}^{\tilde{W}}_{\tilde{\Sigma}}$ will be obtained in the following.

According to the theorem Eq.~\eqref{3.1.25} and following definitions from Eqs.~(\ref{3.1.26}) and (\ref{3.3.1}) ({\em see also Eqs.~(27)-(30) and Definition 3.8 in Ref.~\cite{Bastos}}), a NC Wigner function written in terms of NC variables ${\bf z}$ is given by
\begin{equation}
W^{NC}({\bf z}) = \frac{1}{\sqrt{Det(\mbox{\boldmath$\Omega$})}}
\tilde{W}(\mbox{\boldmath$\xi$}({\bf z}))\equiv\frac{1}{\sqrt{Det(\mbox{\boldmath$\Omega$})}}
\tilde{W}({\bf S}^{-1}{\bf z}),
\end{equation}
from which one can define a NC version of entanglement fidelity given by 
\begin{eqnarray}
\mathcal{F}^{NC}_{\Sigma} &=& (2\pi)^n\, {\hspace{-.1cm}}\int{\hspace{-.15cm}} d^{n}{\bf z}{\hspace{-.1cm}}\int{\hspace{-.15cm}} d^{n}{\bf z}'\, W^{NC}_{in}({\bf z}')\,G^{NC}_{\Sigma}({\bf z} - {\bf z}')\,W^{NC}_{in}({\bf z})\nonumber\\
&=&\frac{(2\pi)^n\,}{{Det(\mbox{\boldmath$\Omega$})}}
{\hspace{-.1cm}}\int{\hspace{-.15cm}} d^{n}\mbox{\boldmath$\xi$}'\bigg{\vert}\frac{\partial {\bf z}'}{\partial\mbox{\boldmath$\xi$}'}\bigg{\vert}
{\hspace{-.1cm}}\int{\hspace{-.15cm}} d^{n}\mbox{\boldmath$\xi$} \bigg{\vert}\frac{\partial {\bf z} }{\partial\mbox{\boldmath$\xi$} }\bigg{\vert}\,\tilde{W}_{in}(\mbox{\boldmath$\xi$}'({\bf z}'))\,
G^{NC}_{\Sigma}\left({\bf z}(\mbox{\boldmath$\xi$}) - {\bf z}'(\mbox{\boldmath$\xi$}'))\right)
\,\tilde{W}_{in}(\mbox{\boldmath$\xi$}({\bf z})).\quad
\label{015}
\end{eqnarray}
Once that the Jacobian determinant $ {\vert}{\partial {\bf z}}/{\partial\mbox{\boldmath$\xi$}}{\vert}$ is given by Eq.~(\ref{Deter}), for
\begin{eqnarray}
G^{NC}_{\Sigma}({\bf z} - {\bf z}') &=& G^{NC}_{\Sigma}\left({\bf z}(\mbox{\boldmath$\xi$}) - {\bf z}'(\mbox{\boldmath$\xi$}'))\right) = 
G^{NC}_{\Sigma}\left({\bf S}(\mbox{\boldmath$\xi$} - \mbox{\boldmath$\xi$}')\right) \nonumber\\&=& \sqrt{\frac{Det(\tilde{\mbox{\boldmath$\Sigma$}})}{Det(\mbox{\boldmath$\Sigma$})}}G_{\tilde{\Sigma}}\left(\mbox{\boldmath$\xi$} - \mbox{\boldmath$\xi$}'\right)= \frac{1}{\sqrt{Det(\mbox{\boldmath$\Omega$})}}G_{\tilde{\Sigma}}\left(\mbox{\boldmath$\xi$} - \mbox{\boldmath$\xi$}'\right), 
\end{eqnarray}
where ${\mbox{\boldmath$\Sigma$}} = {\bf S}\,\tilde{\mbox{\boldmath$\Sigma$}}\,{\bf S}^T$ and ${\mbox{\boldmath$\Omega$}}={\bf S}\,{\bf J}\,{\bf S}^T$, one has
\begin{eqnarray}
\mathcal{F}^{NC}_{\Sigma} &=& (2\pi)^n\, {\hspace{-.1cm}}\int{\hspace{-.15cm}} d^{n}{\bf z}{\hspace{-.1cm}}\int{\hspace{-.15cm}} d^{n}{\bf z}'\, W^{NC}_{in}({\bf z}')\,G^{NC}_{\Sigma}({\bf z} - {\bf z}')\,W^{NC}_{in}({\bf z})\nonumber\\
&=&\frac{(2\pi)^n}{\sqrt{Det(\mbox{\boldmath$\Omega$})}}\,
{\hspace{-.1cm}}\int{\hspace{-.15cm}} d^{n}\mbox{\boldmath$\xi$}
{\hspace{-.1cm}}\int{\hspace{-.15cm}} d^{n}\mbox{\boldmath$\xi$}' \,\tilde{W}_{in}(\mbox{\boldmath$\xi$}')\,
G_{\tilde{\Sigma}}\left(\mbox{\boldmath$\xi$} - \mbox{\boldmath$\xi$}'\right)\,\tilde{W}_{in}(\mbox{\boldmath$\xi$})\nonumber\\&=& \frac{1}{\sqrt{Det(\mbox{\boldmath$\Omega$})}}\mathcal{F}^{\tilde{W}}_{\tilde{\Sigma}},
\label{016}
\end{eqnarray}
which is independent from the SW map.
The above expression for $\mathcal{F}^{NC}_{\Sigma}$ can be straightforwardly specialized to NC Gaussian states identified by $W^{NC}_{in}({\bf z}) = G^{NC}_{\Gamma}({\bf z})$, with $\mbox{\boldmath$\Gamma$}$ arbitrary.
In this case, the expression for the NC version of the fidelity quantifier from Eq.~(\ref{012DD}) can be written as
\begin{eqnarray}
\mathcal{F}^{\Gamma(NC)}_{\Sigma} &=& (2\pi)^n\, {\hspace{-.1cm}}\int{\hspace{-.15cm}} d^{n}{\bf z}{\hspace{-.1cm}}\int{\hspace{-.15cm}} d^{n}{\bf z}'\, G^{NC}_{\Gamma}({\bf z}')\,G^{NC}_{\Sigma}({\bf z} - {\bf z}')\,G^{NC}_{\Gamma}({\bf z})\nonumber\\
&=&\frac{(2\pi)^n}{\sqrt{Det(\mbox{\boldmath$\Omega$})}}\, {\hspace{-.1cm}}\int{\hspace{-.15cm}} d^{n}\mbox{\boldmath$\xi$}{\hspace{-.1cm}}\int{\hspace{-.15cm}} d^{n}\mbox{\boldmath$\xi$}'\, G_{\tilde{\Gamma}}(\mbox{\boldmath$\xi$}')\,G_{\tilde{\Sigma}}({\bf z} - {\bf z}')\,G_{\tilde{\Gamma}}(\mbox{\boldmath$\xi$}')\nonumber\\
&=& \frac{2^n}{\sqrt{Det(\mbox{\boldmath$\Omega$})}\sqrt{{Det[2{\tilde{\mbox{\boldmath$\Gamma$}}}+\tilde{\mbox{{\boldmath$\Sigma$}}}]}}}\nonumber\\
&=& \frac{2^n}{\sqrt{{Det[2{{\mbox{\boldmath$\Gamma$}}}+{\mbox{{\boldmath$\Sigma$}}}]}}} \nonumber\\
&=& \mathcal{F}^{\Gamma}_{\Sigma},
\label{012EE}
\end{eqnarray}
that is, for Gaussian states, the noncommutativity does not affect in any way the fidelity results for continuos variables teleportation protocols: a quantitative result that matches enhanced predictions of the NC phase-space version of the no-cloning theorem.

Turning our attention to more complex quantum states, the above discussion can be extended to the quantum mechanical problem of the $2$-dim NC harmonic oscillator (HO) \cite{Bastos,Rosenbaum}. The quantum Hamiltonian on the NC $x_1-x_2$ plane,
\begin{equation}
\hat{H}^{NC}_{HO}(\hat{\mathbf{z}}) = \hat{\mathbf{z}}\cdot \hat{\mathbf{z}} \leftrightarrow \hat{H}^{NC}_{HO}(\hat{\mathbf{x}},\,\hat{\mathbf{p}}) = \hat{\mathbf{x}}^2+\hat{\mathbf{p}}^2=\sum_{i=1,2} \hat{x}^2_i+ \hat{p}_i^2,
\end{equation}
when re-written in terms of the commutative observables, $\mathit{\hat{q}}_i$ and $\hat{k}_i$, reads 
\begin{equation}
H^W_{HO}(\hat{\mathbf{q}},\,\hat{\mathbf{k}}) = A^2\hat{\mathbf{q}}^2 +B^2\hat{\mathbf{k}}^2 + \gamma \sum_{i,j = 1}^2{\epsilon_{ij}\hat{k}_i \mathit{\hat{q}}_j},
\label{Hamilton}
\end{equation}
where 
\begin{eqnarray}
{A}^2 &\equiv& {\lambda^2 \over 2} + {\eta^2\over 8 \mu^2 \hbar^2}~,\nonumber\\
{B}^2 &\equiv& {\mu^2\over 2} + { \theta^2\over 8 \lambda^2 \hbar^2}~,\nonumber\\
{\gamma} &\equiv& \frac{1}{2\hbar} \left(\theta+\eta\right), 
\label{param}
\end{eqnarray}
and the constraint Eq.~\eqref{constraint} guarantees that $4A^2B^2=1+\gamma^2$.

The {\em stargen}value problem for the Hamiltonian from Eq.~(\ref{Hamilton}) reads\begin{equation}
H^W_{HO} \star W_{n_{\tiny 1},n_{\tiny 2}}(\mbox{\bf \em q},\mathbf{k}) = 
E_{n_{\tiny{1}},n_{\tiny{2}}}\,W_{n_{\tiny{1}},n_{\tiny{2}}} (\mbox{\bf \em q},\mathbf{k}),
\label{help01}
\end{equation}
which, from the analysis developed in Ref.~\cite{Rosenbaum}, results into
\begin{equation}
W_{n_{\tiny{1}},n_{\tiny{2}}} (\mbox{\bf \em q},\mathbf{k}) = \frac{(-1)^{n_1+n_2}}{\pi^2\hbar^{2}}\exp\left[{-\left(\Omega_{+}+\Omega_{-}\right)/\hbar}\right] \, L^0_{n_1} \left(\Omega_{+}/\hbar\right) \,L^0_{n_2}\left(\Omega_{-}/\hbar\right),
\label{Lague01}
\end{equation}
where $L^0_n$ are the associated Laguerre polynomials, $n_1$ and $n_2$ are non-negative integers, and
\begin{equation}
{\Omega}_{\pm} = {A\over B}\mbox{\bf \em q}^2 + {B\over A}\mathbf{k}^2 \mp 2 \sum_{i,j = 1}^2{\left(\epsilon_{ij}k_i \mathit{q}_j\right)},
\end{equation}
such that the energy spectrum is given by
\begin{equation}
E_{n_{\tiny 1},n_{\tiny 2}} = \hbar\left[2\alpha\beta(n_1 + n_2 + 1) + \gamma (n_1 - n_2)\right],
\end{equation}
and one has
\begin{equation}
\int^{^{+\infty}}_{_{-\infty}}\hspace{-.5cm}d\mathit{q}_1\int^{^{+\infty}}_{_{-\infty}}\hspace{-.5cm} dk_1\int^{^{+\infty}}_{_{-\infty}}\hspace{-.5cm} d\mathit{q}_2\int^{^{+\infty}}_{_{-\infty}}\hspace{-.5cm} dk_2 \,\,\rho_{n_{\tiny{1}},n_{\tiny{2}}}^W (\mbox{\bf \em q},\mathbf{k}) = 1.
\end{equation}

The mapped variables ${\Omega}_{\pm}$ can be straightforwardly identified with $\Omega_{\pm} = |\alpha_1 \pm i \alpha_2|^2$ for
\begin{equation}
\alpha_j = x_j + ip_j,\quad j=1,\,2,
\end{equation}
which shows that {input} NC Wigner functions described by Eq.~\eqref{Lague01} in the $\Omega_{+}-\Omega_{-}$ plane are equivalent to the ordinary commutative Wigner functions with $\Omega_{\pm} = |\alpha_1 \pm i \alpha_2|^2$
given in terms of
\begin{equation}
\alpha_j = \sqrt{\frac{A}{B}}q_j + i\sqrt{\frac{B}{A}}k_j,\quad j=1,\,2,
\end{equation}
which does not change the phase-space volume of integration and keeps valid the general result Eq.~(\ref{014B}).
By considering the Gaussian channel described by Eq.~\eqref{Wigner2}, NC and ordinary QM yield the same {\em output} states and, therefore, entanglement fidelity results are not affected.

\section{Quantum teleportation}

Quantum teleportation is a purely quantum phenomenon first discussed by Bennet {\em et al.} \cite{Bennett1993} for discrete variable quantum systems. It allows for the transmission of a single particle quantum state from one location to another via the exchange of two classical information bits making use of the quantum entanglement. Due to these properties and from the fact that the state of the teleported particle at the {\em input} location is destroyed, there is neither violation of causality nor violation of the no-cloning theorem. The generalization of this result for continuous variables was first put into place by Vaidman \cite{Vaidman1994}, in a setup similar to the one of discrete variables. In the following analysis, the teleportation process is discussed in the Wigner formalism, first for the usual QM and then in the NC framework. 

\subsection{Ideal teleportation -- Standard QM case}

In the first step in the preparation of this process, the two intervening parts must share an entangled state of two particles, $W^{\mathrm{EPR}}\left(x^A,p^A,x^B,p^B\right)$, where $x^{A,B}$, $p^{A,B}$ correspond to the position and momentum of particles belonging to Alice and Bob, respectively. Let the teleported state be described by the Wigner function $W^{in}(x^C,p^C)$. Now one of the intervening parts, Alice, gets particles $A$ and $C$ and Bob gets particle $B$. The complete state of the system composed by the three particles is given by:
\begin{equation}
W(x^A,p^A,x^B,p^B,x^C,p^C)=W^{\mathrm{EPR}}\left(x^A,p^A,x^B,p^B\right)W^{in}(x^C,p^C).
\end{equation}
Alice now uses a beam splitter on particles $A$ and $C$ so that their positions and momenta can be written as:
\begin{equation}
x_\pm=\frac{1}{\sqrt{2}}(x^A\pm x^C), \qquad p_\pm=\frac{1}{\sqrt{2}}(p^A\pm p^C),
\end{equation}
which, once introduced into the Wigner function argument yields:
\begin{equation} \label{Wigner_Comm_pm}
W=W^{\mathrm{EPR}}\left(\frac{1}{\sqrt{2}}(x_++x_-),\frac{1}{\sqrt{2}}(p_++p_-),x^B,p^B\right)W^{in}\left(\frac{1}{\sqrt{2}}(x_+-x_-),\frac{1}{\sqrt{2}}(p_+-p_-)\right).
\end{equation}
Since $\left[x_+,p_-\right]=0$, Alice can now make a measurement of these two quantities. Then, the final state of particle $B$ is given by:
\begin{eqnarray}
W^{out}(x^B,p^B)&=&W(x^B,p^B\, | \,x_+,p_-)\nonumber\\&=&{\hspace{-.1cm}}\int{\hspace{-.15cm}} dx_-{\hspace{-.1cm}}\int{\hspace{-.15cm}} dp_+\,W^{\mathrm{EPR}}\left(\frac{1}{\sqrt{2}}(x_++x_-),\frac{1}{\sqrt{2}}(p_++p_-),x^B,p^B\right)\nonumber\\&&
\qquad \qquad\qquad\qquad \qquad \qquad \times W^{in}\left(\frac{1}{\sqrt{2}}(x_+-x_-),\frac{1}{\sqrt{2}}(p_+-p_-)\right).
\end{eqnarray}\normalsize
If the shared entangled state of particles $A$ and $B$ is a maximally entangled state, i.e. $W^{EPR}=\delta(x^A+x^B)\delta(p^A-p^B)$, then the final state of particle $B$ is:
\begin{equation}
W^{out}(x^B,p^B)=W^{in}\left(x^B+\sqrt{2}x_+,p^B-\sqrt{2}p_+\right).
\end{equation}
The next step in the teleportation process is the communication of Alice results to Bob. This corresponds to the value of $x_+$ and $p_-$ and these are usually exchanged via classical communication. Then Bob proceeds with the adequacy of his state according to the information received, adjusting the position and momentum of particle $B$, i.e. $x^B+\sqrt{2}x_+\rightarrow x^B$ and $p^B-\sqrt{2}p_+\rightarrow p^B$, which leads to:
\begin{equation}
W^{out}(x^B,p^B)=W^{in}(x^B,p^B).
\end{equation}
The above condition implies that the final state of particle $B$ is the same as the initial state of particle $C$. Therefore, for any $W^{in}$, the particle $B$ can be put into that state by the procedure described above. Since the particle $B$ does not need to be in the same physical location as the other two particles, the state was teleported to its location.

This procedure can be straightforwardly generalized to more than one dimension as long as an EPR state can be prepared. This implies that $\left[x^i+x^j,p^i-p^j\right]=0$ for $i\neq j$ and $i,j=A,\,B,...$.



\subsection{Ideal teleportation -- NC QM case}

For the NC scenario one should address to more than one dimension teleportation protocols. One then considers the same previous setup for three particles, now with position $(x^i_1,x^i_2) \equiv (x^i,y^i)$ and momentum $(p_{1}^i,p_{2}^i)\equiv(p_{x}^i,p_{y}^i)$, with $i,j=A,\,B,\, C$. The three particle Wigner function is then:
\begin{equation}
W_{NC}=W^{EPR}_{NC}(x^A,y^A,p_{x}^A,p_{y}^A,x^B,y^B,p_{x}^B,p_{y}^B)W^{in}_{NC}(x^C,y^C,p_{x}^C,p_{y}^C).
\end{equation}
Using a beam splitter, in the same way as in the previous case, but in $2$-dim, we introduce a new set of variables:
\begin{equation}
x_\pm=\frac{1}{\sqrt{2}}(x^A\pm x^C), \qquad y_\pm=\frac{1}{\sqrt{2}}(y^A\pm y^C), \qquad p_{x_\pm}=\frac{1}{\sqrt{2}}(p_{x}^A\pm p_{x}^C), \qquad p_{y_\pm}=\frac{1}{\sqrt{2}}(p_{y}^A\pm p_{y}^C),
\end{equation}
which can be used to rewrite the Wigner function as in Eq.~\eqref{Wigner_Comm_pm}. The NC issues impose some constraints upon the quantities measured by Alice: now $x_+$ and $y_+$ cannot be measured simultaneously, since $\left[x_+,y_+\right]=2\mathrm{i}\theta\neq 0$. Nevertheless, it is possible to choose a set of four observables to measure simultaneously in a way that the final state of particle $B$ matches the one of the initial particle, $C$. These observables are $x_+$, $y_-$, $p_{x_-}$ and $p_{y_+}$, which constitute a complete set of commuting observables. The Wigner function describing the state of particle $B$ after Alice's measurement is then given by:
\begin{equation}
W_{NC}^{out}(x^B,y^B,p_{x}^B,p_{y}^B)={\hspace{-.1cm}}\int{\hspace{-.15cm}} dx_-{\hspace{-.1cm}}\int{\hspace{-.15cm}} dy_+{\hspace{-.1cm}}\int{\hspace{-.15cm}} dp_{x_+}{\hspace{-.1cm}}\int{\hspace{-.15cm}} dp_{y_-} W_{NC}^{EPR}\,W^{in}_{NC}.
\end{equation}
If, as in the ordinary QM teleportation, one assumes the shared state to be maximally entangled, i.e.
\begin{equation}
W_{NC}^{EPR}=\delta(x^A+x^B)\delta(y^A-y^B)\delta(p_{x}^A-p_{x}^B)\delta(p_{y}^A+p_{y}^B),
\end{equation}
then the final state of particle $B$ would be:
\begin{equation}
W^{out}_{NC}=W^{in}_{NC}(x^B+\sqrt{2}x_+,y^B-\sqrt{2}y_-,p_{x}^B-\sqrt{p_{x_-}},p_{y}^B+\sqrt{2}p_{y_+}).
\end{equation}
When Alice communicates her results on the measurements of $x_+$, $y_-$, $p_{x_-}$ and $p_{y_+}$, Bob can make the necessary unitary transformations on particle $B$ so to have:
\begin{equation}
W^{out}_{NC}=W^{in}_{NC}(x^B,y^B,p_{x}^B,p_{y}^B),
\end{equation}
successfully teleporting the unknown original state of particle $C$ into particle $B$. 

Therefore, using the above theoretical setup, quantum teleportation of particle states with $F=1$ is also possible in the framework of NC QM. Of course, the shared state used in this procedure is not experimentally viable and the results for the teleportation probabilities shall be also constrained by an EPR state as given by the preliminaries of Sec. IV,  i.e. when squeezed Gaussian states are used as shared entangled states.

\section{Conclusions}

The fundamental tenets related to the reproducibility of quantum states through quantum cloning and quantum teleportation were examined in the framework of phase-space NC QM. The no-cloning theorem was cast into the Wigner formalism for QM and then generalized for phase-space NC QM, showing that a perfect copy of a state is impossible to be achieved within the NC framework as well. It is seen that the only feature required for this result to hold is the unitarity, which is also a feature of NC QM.

In what concerns quantifiers of the above mentioned quantum processes, the quantum fidelity for continuous variable teleportation protocols in the NC phase-space was computed using the NC Wigner function definitions (cf. Ref.~\cite{Bastos}).
It has been shown that, when computed for $2$-dim Gaussian and HO \textit{input} states, the noncommutativity does not affect the fidelity results of duplicated states. In fact, the results for quantum fidelity are shown to be independent of the SW map which relates NC and ordinary quantum systems. This is in line with results of Ref.~\cite{Bastos}, where it is shown that physical results such as expectation values and transition amplitudes are independent of the SW map.

Finally, a procedure for quantum teleportation of continuous variables in the phase-space NC QM was developed as to take into account some subtle modifications on the commutative teleportation protocol. These modifications amounted to change the preparation of the entangled state in order to account for the additional NC relations so to agree with the uncertainty principles imposed by the deformed HW algebra. To summarize, NC QM does impose, likewise in QM, relevant obstacles to the reproduction of states as well as associated duplication protocols.

\vspace{.5 cm}
{\em Acknowledgments} -- The work of PL is supported by FCT (Portuguese Funda\c{c}\~ao para Ci\^encia e Tecnologia) grant PD/BD/135005/2017. The work of AEB is supported by the Brazilian agency FAPESP (grant 17/02294-2). The work of OB is partially supported by the COST action MP1405.

\end{document}